\documentclass[
  ,final            
  ]
  {aipproc}

\layoutstyle{8x11single}

\begin{document}

\def\Swift{\emph{Swift}}
\newcommand{\lesssim}{\lower.5ex\hbox{$\; \buildrel < \over\sim \;$}}
\newcommand{\gtrsim}{\lower.5ex\hbox{$\; \buildrel > \over\sim \;$}}
\newcommand{\apj}{{\it Astrophysical J.,~}}

\title{Gamma Ray Bursts in the Swift and GLAST Era}


\author{Truong Le and Charles D. Dermer}{
  address={Code 7653, Space Science Division,
Naval Research Laboratory\break Washington, DC 20375, USA} }



\begin{abstract}
We summarize our model for long-duration gamma ray bursts (GRBs) that
fits the redshift ($z$) distributions measured with Swift and missions
before Swift, and the pre-Swift GRB jet opening-angle distribution inferred from
achromatic breaks in the optical light curves. We
find that the comoving rate density of GRB sources exhibits positive
evolution to $z \gtrsim 3$ -- 5, whereas the star formation rate
inferred from measurements of the blue and UV luminosity density peaks
at $z\sim 1$ -- 3.  The mean intrinsic beaming factor of GRBs is found
to be $\approx 34$ -- 42, and we predict that the mean GRB optical jet opening
half-angle measured with Swift is $\approx 10^\circ$. We estimate the
number of GRBs per year that GLAST is expected to observe based on
ratios of BATSE and EGRET GRB fluences.
\end{abstract}

\maketitle


\section{Introduction}

With the launch of the \Swift~satellite~\citep{geh04}, rapid follow-up
studies of GRBs triggered by the Burst Alert Telescope (BAT) on
\Swift~became possible. A fainter and more distant population of GRBs
than found with the pre-\Swift~satellites CGRO-BATSE, BeppoSAX,
INTEGRAL, and HETE-2 is detected~\citep{ber05}. The mean redshift of
41 pre-\Swift~GRBs that also have measured optical afterglow
 beaming breaks~\citep{fb05}
is $\langle z\rangle = 1.5$, while 16 GRBs discovered by \Swift~have
$\langle z\rangle = 2.72$~\citep{jak06}. The null hypothesis that the
redshift distributions of \Swift~and pre-\Swift~GRBs are the same
is rejected in several statistical tests at the  $\gtrsim$97\%
confidence level \citep{bag06}.

In recent work \citep{ld06}, we considered whether the differences
between the pre-\Swift~and \Swift~redshift distributions can be
explained with a physical model for GRBs that takes into account the
different flux thresholds of GRB detectors, where we assume the
detector threshold for \Swift~and pre-\Swift~GRBs to be $\sim 10^{-8}$
and $\sim 10^{-7}\ \rm ergs \ cm^{-2} \ s^{-1}$, respectively
\citep{ld06}, corresponding to the minimum peak fluxes of GRBs with
redshifts measured by Swift and missions before Swift. We summarize
\citep{ld06} and make an independent estimate of the number of GRBs that
GLAST should detect.

\section{Results}

We parameterize the distribution of GRB jet opening angles
 with a function of the form $dN/d\mu_j \propto
(1-\mu_j)^s$, where arccos$(\mu_j)$ is the jet half-angle, and find
best fit values for the $\gamma$-ray energy release ${\cal E}_\gamma$
for different functional forms of the comoving rate density of GRBs
(see Fig.~1, left panel). We simplify the analysis with a flat GRB
$\nu F_\nu$ spectrum, and assume that the properties of GRBs do not
change with time. Adopting the uniform jet model where the energy per
solid angle is roughly constant throughout the GRB jet, and taking a
mean intrinsic duration of 10 seconds, we obtain best-fit values with
a corrected emitted $\gamma$-ray energy ${\cal E}_\gamma = 4 \times
10^{51}$ ergs, jet opening angles in the range $0.05$ -- $0.7$
radians, and with $s\approx -1.25$.

Our analysis shows that good fits to the pre-\Swift~and
\Swift~redshift and opening angle data require a GRB rate
history that rises faster than the star formation rate (SFR2,3,4)
at high redshifts (SFR5 and SFR6; see Fig.~1, left panel).
\begin{figure}[t]
\resizebox{1\textwidth}{!} {\includegraphics[width=7.5in]{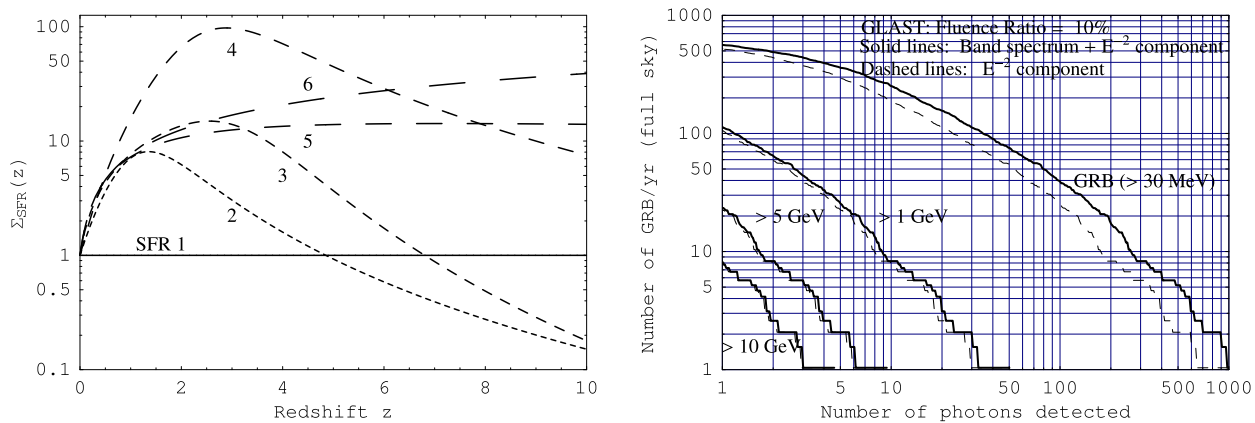}}
\caption{{\it Left panel:} star formation rate (SFR) functions
used in our GRB study. The solid line (SFR1) is a constant
comoving density; SFR2 and SFR4 are low and high ranges for the
SFR; SFR3 is a recent fit \citep{hb06} to the data; SFR5 and SFR6
are GRB rate histories that give a good fit to the \Swift~and
pre-\Swift~redshift distribution and the pre-Swift opening angle
distribution \citep{ld06}. {\it Right panel:} Estimate of the
number of GRBs per year GLAST will observe, based on measured
EGRET/BATSE fluence ratios.} %

\end{figure}
The results of our fitting indicate that GRB activity was greater in
the past and is not simply proportional to the bulk of the star
formation as traced by the blue and UV luminosity density of the
universe. Furthermore, our model predicts the mean intrinsic beaming
factor of GRB jetted outflows contributing to the optical breaks
 to be in the range from $\approx 34-42$. The analysis
also indicates that the average jet opening half-angle of GRBs
detected with Swift is $\langle\theta_j\rangle \approx 10^\circ$,
compared to the pre-\Swift~average of $7^\circ$ \citep{ld06}. Thus we
expect to detect more faint low-redshift, large opening angle GRBs
that pre-\Swift~satellites could not detect (see also \citep{psf03}).

We also estimate the number of GRBs per year GLAST will observe using
the fluence ratios of BATSE ($\simeq 20$ keV -- 2 MeV) and EGRET (100
MeV -- 5 GeV) for the 5 BATSE bursts also detected in the EGRET spark
chamber~\citep{din95}. We find that the average BATSE/EGRET fluence
ratio is $\approx 5$\%, though with large scatter and not taking into
account Earth occultation (e.g., for GRB 940217) \citep{der05},
deadtime effects \citep{din95}, and anomalous GRBs like GRB 941017
\citep{gon03}. Our estimate uses the BATSE 4B fluence distribution, a
Band function with Band $\beta = -2.5$ added to a $>$ 30 MeV spectral
component with a $-2$ photon index in the EGRET/GLAST
range~~\citep{dcs98}, and power-law approximations for the EGRET and GLAST
 effective areas \citep{dd05}.

Fig.1, right panel, gives the resulting estimate of the number of
GRBs GLAST should observe per year full-sky for a fluence ratio of
10\%, for different integral photon numbers and energies.  We find
that there will be $\approx 350$ ($\approx 25$) GRBs/yr full-sky
from which the GLAST LAT would detect $\geq 5$ photons with energy
$E> 30$ MeV ($> 1$ GeV), and $\approx 25$ ($\approx 8$) GRBs/yr
full-sky with $\geq 1$ photon with $E>5$ GeV ($>10$ GeV). A
similar conclusion \citep{omo06}, obtained by extrapolating BATSE
results fitted with the Band function to LAT energies, gives an
estimate of 50-70 GRBs/yr with $>5$ photons with $E> 30$ MeV. Our
analysis also shows that GRBs give very little ($\lesssim 1$\% )
contribution to the diffuse extragalactic $\gamma$-ray background.
\begin{theacknowledgments}
T.~L. is funded through NASA {\it GLAST} Science Investigation
No.~DPR-S-1563-Y. The work of C.~D.~D. is supported by the Office
of Naval Research. We thank Nicola Omodei and Brenda Dingus for
discussions, and Brenda Dingus for pointing out how deadtime
effects could be important for EGRET.
\end{theacknowledgments}

\end{document}